# Measuring the relatedness between scientific publications using controlled vocabularies

**Emil Dolmer Alnor**

0000-0002-8536-9442
*ea@ps.au.dk*
Danish Centre for Studies in Research and Research Policy, Aarhus University, Denmark

**Abstract**
Measuring the relatedness between scientific publications is essential in many areas of bibliometrics and science policy. Controlled vocabularies provide a promising basis for measuring relatedness and are widely used in combination with Salton's cosine similarity. The latter is problematic because it only considers exact matches between terms. This article introduces two alternative methods—soft cosine and maximum term similarities—that account for the semantic similarity between non-matching terms. The article compares the accuracy of all three methods using the assignment of publications to topics in the TREC 2006 Genomics Track and the assumption that accurate relatedness measures should assign high relatedness scores to publication pairs within the same topic and low scores to pairs from separate topics. Results show that soft cosine is the most accurate method, while the most widely used version of Salton's cosine is markedly less accurate than the other methods tested. These findings have implications for how controlled vocabularies should be used to measure relatedness.

## Introduction

Measuring the relatedness between scientific publications has important applications in many areas of bibliometrics and science policy. It is central in measuring interdisciplinarity (Rafols & Meyer, 2010; Stirling, 2007), aiding information retrieval (Lin & Wilbur, 2007), clustering science into topics (Waltman & Van Eck, 2012), and analyzing the causes and consequences of researchers adapting their research trajectories (Hill et al., 2025; Myers, 2020). The relatedness between scientific publications is often measured using their citation relation, textual similarity, or a combination of the two (Waltman et al., 2020). While applicable to many publications, these approaches have some known issues, e.g., citation time lags, synonyms, homonyms, and polysemy (Glänzel & Thijs, 2017, pp. 1072-1073).

Using controlled vocabularies to measure relatedness addresses these issues because publications are often indexed with terms shortly after publication and because controlled vocabularies provide consistency and uniqueness in how terms refer to concepts (Hersh, 2020). Controlled vocabularies are used to measure publication relatedness in numerous studies in diverse areas of science, such as physics (Jia et al., 2017; Yu et al., 2021), computer science (Salatino et al., 2023), and biomedical sciences (Ahlgren et al., 2020; Boudreau et al., 2016; Colliander & Ahlgren, 2019; de Rassenfosse et al., 2022).

All studies cited above rely on Salton's cosine similarity (Salton & McGill, 1983). The idea behind Salton's cosine is that the more terms two publications have in common, the more related they are. Salton's cosine is intuitive, has a simple mathematical formulation, and is fast to compute. It has a major conceptual problem, however, in that it only counts exact matches between terms. To illustrate this problem, consider three publications, $p_a$, $p_b$, and $p_c$, which are all indexed with the term "Health Personnel" but differ in their second term: $p_a$ has "Heart Arrest", $p_b$ has "Heart Failure", while $p_c$ has "Horse Diseases". Intuitively, $p_a$ and $p_b$ should be more related than $p_a$ and $p_c$ because "Heart Arrest" is semantically more similar to "Heart Failure" than to "Horse Diseases". However, because Salton's cosine only considers exact matches, all three publications will be equally related to each other. Clearly, this does not align with the intuitive degree of relatedness between the publications.

The aims of this article are 1) to develop methods that account for the semantic similarity between non-matching terms when using controlled vocabularies to measure the relatedness between scientific publications, and 2) to test which method provides the most accurate measure.

The article proceeds as follows. The next section provides a brief introduction to controlled vocabularies and the controlled vocabulary used in this article, namely Medical Subject Headings (MeSH). Section three introduces Salton's cosine in more detail, its extension proposed by Ahlgren et al. (2020), and two alternative methods—soft cosine and maximum term similarities—that account for the semantic similarity between non-matching terms. Section four describes how the assignment of publications to topics in the TREC 2006 Genomics Track is used to compare the accuracy of the different methods, using the assumption that an accurate relatedness measure should assign high relatedness to publication pairs within the same topic and low relatedness to publication pairs from separate topics. Section five shows that soft cosine is the most accurate method, while the most widely used version of Salton's cosine is markedly less accurate than the other methods tested. The final section discusses the implications of these results for how controlled vocabularies should be used to measure relatedness.

## Controlled vocabularies

Hersh (2020, p. 184) defines two important terms in understanding controlled vocabularies: "A *concept* is an idea or object that occurs in the world, such as the condition under which human blood pressure is elevated. A *term* is the actual string of one or more words that represent a concept, such as Hypertension or High Blood Pressure". Controlled vocabularies solve the issues of synonymy and homonymy by having a 1:1 mapping between concepts and terms (each concept is always represented by the same term).

The controlled vocabulary used in the empirical parts of this article is the MeSH thesaurus. MeSH contains more than 30,000 terms that describe scientific concepts and that are hierarchically organized (U.S. National Library of Medicine, 2024). For example, "Heart Arrest" is a narrower term to "Heart Diseases". Terms can either be assigned to a publication as major, indicating that they reflect the primary topics or concepts of the publication, or minor, indicating that they reflect concepts of lesser importance to the publication. The publications analyzed in this article are indexed with 15 MeSH terms on average and with four major terms on average. Around 90% of terms have qualifiers describing the specific aspect of the concept. For example, "rehabilitation" can be used as a qualifier for "Heart Arrest". Until 2022, MeSH indexing was done by human field experts, and since April 2022 human field experts have curated and reviewed machine-based indexing.

The methods developed in this article could be applied to any controlled vocabulary, thesaurus, or ontology as long as it describes the hierarchical relationship between its terms, e.g., the IEEE Thesaurus, the Physics Subject Headings (PySH), the Physics and Astronomy Classification Scheme (PACS), the Computer Science Ontology (CSO), or the ACM Computing Classification System (CCS).

## Measuring the relatedness between scientific publications using controlled vocabularies

### Salton's cosine similarity

The intuition of Salton's cosine is that the more terms two publications have in common relative to their total number of terms, the more related they are. Formally, let publication $p_a$ be indexed with a set of terms, $P_a$. Represent $p_a$ with vector $\mathbf{a} = [a_1, \ldots, a_n]$, where $n$ is the number of terms in the controlled vocabulary and $a_i = \mathrm{I}(t_i \in P_a)$. The indicator function $\mathrm{I}(\cdot)$ evaluates to 1 when its contents are true—here, when $p_a$ is indexed with term $i$. Using Salton's cosine, the relatedness between two publications $p_a$ and $p_b$ is

$$\mathrm{rel}_{Salton's\ cosine}(p_a, p_b) = \frac{\sum_{i=1}^{n} a_i b_i}{\sqrt{\sum_{i=1}^{n} a_i^2} \sqrt{\sum_{i=1}^{n} b_i^2}} = \frac{\mathbf{a} \cdot \mathbf{b}}{\|\mathbf{a}\| \|\mathbf{b}\|} \quad (1)$$

While this should capture some of the relatedness between $p_a$ and $p_b$, there are some deficiencies: two publications sharing a very frequent term are deemed as related as two publications sharing a very rare term, and the method utilizes neither the distinction between major and minor terms nor their qualifiers when using MeSH.

Colliander and Ahlgren (2019) introduce a vector representation of $p_a$ that addresses these deficiencies. The intuition is that the more information two publications share relative to their total information, the more related they are. To quantify this, we define the information content (IC) for terms, which is low for common terms—that is, when the frequency of the term and/or its descendants is high. Formally, let $\#(t_i)$ denote the frequency of term $i$ in the dataset, let $D_i$ denote the set of terms that are descendants to $t_i$ either directly or indirectly, and let $n$ denote the number of unique terms. Then, the IC of $t_i$ is defined as

$$\mathrm{IC}(t_i) = -\log \left( \frac{\#(t_i) + \sum_{d \in D_i} \#(d)}{\sum_{j=1}^{n} \#(t_j) + \sum_{d \in D_j} \#(d)} \right) \quad (2)$$

Now, we utilize the distinction between major and minor terms and the IC of terms by representing $p_a$ with vector $\grave{\mathbf{a}} = [\grave{a}_1, \ldots, \grave{a}_n]$ and by letting

$$\grave{a}_i = \begin{cases} 0 & \text{if } t_i \notin P_a \\ \text{IC}(t_i) & \text{if } t_i \text{ is a minor term in } P_a \\ \text{IC}(t_i) \times w & \text{if } t_i \text{ is a major term in } P_a \end{cases} \quad (3)$$

where $w$ is the weight assigned to major terms. The empirical part of the article tests major weights from 2 to 10.

To utilize the qualifiers, we introduce a third vector representation of $p_a$ that has length $n + n \times q$, where $q$ denotes the number of unique qualifiers. This means that there is a position for each unique term and for each unique term-qualifier combination. Specifically, with $t_{i_j}$ denoting term $i$ with qualifier $j$, represent publication $p_a$ by vector $\breve{\mathbf{a}} = [\breve{a}_1, \breve{a}_{1_1}, \ldots, \breve{a}_{1_q}, \breve{a}_2, \breve{a}_{2_1}, \ldots \breve{a}_{2_q}, \ldots, \breve{a}_{n_q}]$. To utilize the qualifiers, let $\breve{a}_{i_j} = \text{I}\left(t_{i_j} \in P_a\right)$ and let $\breve{a}_i = \grave{a}_i$. Note that a term can appear with multiple qualifiers in a publication and that many term-qualifier combinations are nonsensical and will never exist in practice (Colliander & Ahlgren, 2019, p. 292).

The empirical part of this article compares the accuracy of the three vector representations $\mathbf{a}$, $\grave{\mathbf{a}}$, and $\breve{\mathbf{a}}$. While $\grave{\mathbf{a}}$ and $\breve{\mathbf{a}}$ address the deficiencies mentioned above, using them in Salton's cosine still only considers exact matches between terms. To address this, we first introduce two methods for measuring the semantic similarity between terms and then two methods for using the similarity in measuring publication relatedness.

**Measuring term similarity**

To measure term similarity, we construct a graph with terms as nodes and edges between child-parent pairs of terms. Formally, let $G_1 = (V, E)$ be an unweighted undirected graph, where $V = \{t_1, \ldots, t_n\}$ and $E = \{e_1, \ldots e_m\}$, where $m$ is the number of parent-child relationships between the terms. Specifically, let $\{t_i, t_j\} \in E$ if $t_i$ is the child or parent of $t_j$. The distance between two terms, $\text{dist}(t_i, t_j)$, is defined as the number of edges in the shortest path between them. Following Zhu et al. (2009), we convert distance to similarity using

$$\text{sim}(t_i, t_j) = e^{-\left(\frac{\text{dist}(t_i, t_j)}{\lambda}\right)} \quad (4)$$

Increasing $\lambda$ makes distant terms more similar. The empirical part of this article tests which value of $\lambda$ yields the most accurate relatedness measure.

Since the graph is unweighted, the distance between all parent-child pairs of terms is equal. If a term has a single child with (almost) equal IC as itself, it could be argued that the distance between this parent-child pair is shorter than the distance between a parent-child pair with a large difference in IC. To test this argument, we introduce a second method of defining term distance. Let $G_{\Delta IC} = (V, E, w)$ be a weighted undirected graph identical to $G_1$ except that each edge has a weight:

$$w(t_i, t_j) = \Delta IC(t_i, t_j) = |IC(t_i) - IC(t_j)| \tag{5}$$

In words, the distance between two directly related terms is their difference in IC, and the distance between any two terms is the sum of weights of the edges in the shortest weighted path between them. Distance is again converted to similarity using equation 4.

**Using term similarity to measure publication relatedness**

We now describe the methods for using term semantic similarity to measure publication relatedness.

**Soft cosine**

The first approach is an extension of Salton's cosine. Whereas Salton's cosine quantifies the extent to which terms are the same, this approach quantifies the extent to which they are similar. The approach has been independently introduced by Zhou et al. (2012) as "similarity-weighted cosine measure" and Sidorov et al. (2014) as "soft cosine". We call it soft cosine because this name has gained more traction; e.g., a software implementation using this name is available in the Gensim-package (Řehůřek & Sojka, 2010).

Let $\underset{n \times n}{S}$ be a matrix with $s_{ij} = \text{sim}(t_i, t_j)$ measuring the similarity between terms $i$ and $j$. In general, $\text{sim}(\cdot)$ can be any similarity measure, but here it is natural to use equation 4. The empirical part of the article tests equation 4 using both the unweighted and weighted graphs. Recall that publications $p_a$ and $p_b$ can be represented with term vectors **a** and **b**. Then, the soft cosine similarity between $p_a$ and $p_b$ is

$$\text{rel}_{soft\ cosine}(p_a, p_b) = \frac{\sum \sum_{i,j=1}^{n} s_{ij} a_i b_j}{\sqrt{\sum \sum_{i,j=1}^{n} s_{ij} a_i a_j} \sqrt{\sum \sum_{i,j=1}^{n} s_{ij} b_i b_j}} = \frac{\mathbf{a}^T \mathbf{S} \mathbf{b}}{\sqrt{\mathbf{a}^T \mathbf{S} \mathbf{a}} \sqrt{\mathbf{b}^T \mathbf{S} \mathbf{b}}} \tag{6}$$

The empirical part of the article compares the accuracy of using the non-weighted (**a**) and IC and major weighted (**à**) term vectors in equation 6. Qualifiers (**ă**) are not used because there is no obvious method to model similarity between term-qualifier pairs. To understand how soft cosine relates to Salton's cosine, note that if there is no similarity between terms ($s_{ij} = 0$ for $i \neq j$), soft cosine and Salton's cosine are equivalent. In this regard, Salton's cosine can be understood as a special case of soft cosine similarity with $\mathbf{S} = \mathbf{I}$, or soft cosine similarity can be understood as Salton's cosine similarity using an oblique (non-orthogonal) basis.

**Maximum term similarities**

In the second approach, the relatedness between publications is a function of the maximum similarities between their terms. Recall that $P_a$ and $P_b$ denote the set of terms in publications $p_a$ and $p_b$, respectively. Following Zhu et al. (2009), we then define the relatedness between $p_a$ and $p_b$ as

$$\text{rel}_{mts_1}(p_a, p_b) = \frac{\sum_{t_i \in P_a} \max_{t_j \in P_b} \text{sim}(t_i, t_j) + \sum_{t_j \in P_b} \max_{t_i \in P_a} \text{sim}(t_j, t_i)}{|P_a| + |P_b|} \quad (7)$$

That is, it is the sum of maximum similarities between terms in $P_a$ and terms in $P_b$ and vice versa, divided by the number of terms in $P_a$ and $P_b$. In this approach, minor and major terms contribute equally to computing publication relatedness, which is not desirable. To address this problem, Zhu et al. (2009) propose the "slim" method, which excludes minor terms from $P_a$ and $P_b$. However, doing this discards the information inherent in the minor terms. Instead, we propose applying a weight, $w$, to major terms:

$$\text{rel}_{mts_w}(p_a, p_b) = \frac{\sum_{t_i \in P_a} \max_{t_j \in P_b} \text{sim}(t_i, t_j) \times w_{t_i} + \sum_{t_j \in P_b} \max_{t_i \in P_a} \text{sim}(t_j, t_i) \times w_{t_j}}{\sum_{t_i \in P_a} w_{t_i} + \sum_{t_j \in P_b} w_{t_j}} \quad (8)$$

That is, the relatedness between $p_a$ and $p_b$ is the sum of their weighted maximum term similarities divided by the sum of weights. Minor terms receive weight 1, and the article evaluates major weight from 2 to 20, e.g., $w_{t_i} = 2$ if $t_i$ is a major term in $p_a$ and 1 otherwise. Note that equations 7 and 8 yield identical results when $w_{t_i} = w_{t_j} = 1$ for both major and minor terms.

## Evaluating relatedness measures

### Limitations of the standard approach

The tests used in this article to evaluate the relatedness measures differ from the typical test, which uses the relatedness measures in a clustering algorithm and then compares the retrieved topics to some golden standard of topics (Ahlgren et al., 2020; Boyack & Klavans, 2020; Klavans & Boyack, 2017; Waltman et al., 2020; Zhu et al., 2009). The reason for this deviation is that the typical test only indirectly evaluates the relatedness measures themselves. Clustering solutions are produced by both the relatedness measure used in the similarity matrix and the algorithm used to cluster the similarity matrix (Held, 2022; Šubelj et al., 2016). Therefore, the accuracy of a clustering solution cannot be attributed solely to the relatedness measure. While using different relatedness measures in a clustering algorithm and assessing the accuracy of the resulting clusters provides insight into the accuracy of those relatedness measures within that specific clustering algorithm, it provides less insight into their accuracy within other clustering algorithms, and even less insight into the accuracy of the relatedness measures themselves. As highlighted in the introduction, relatedness measures have many applications beyond clustering.

**TREC Genomics**

The benchmark data comes from the TREC Genomics Track 2006 (Hersh et al., 2006). Here, teams of information scientists were given 28 information needs (referred to as topics below) and the task to nominate relevant text passages from a set of approximately 160,000 publications. All topics concerned the role of genes in biological processes or diseases, such as "What is the role of PrnP in mad cow disease?" Each of the 92 teams submitted 1,000 text passages spread across the 28 topics. Subsequently, field experts evaluated these passages as either "not relevant", "possibly relevant", or "relevant" to the topic. The relevance of a publication to a topic is here defined as the maximum relevance of any of its passages to that topic. The tests below only include topics where at least 10% of the nominated publications are relevant or possibly relevant and only includes relevance judgements of "not relevant" and "relevant". The final dataset contains 3,841 relevance judgements (26.4% relevant) spread across 3,226 publications and nine topics.

**Benchmark test 1**

The first test evaluates the ability of the relatedness measures to compute higher relatedness between pairs of publications both deemed relevant to the same topic (same-topic pairs) compared to pairs where one publication is relevant to a topic and the other is not (separate-topic pairs). Using a medical analogy, this test resembles measuring differences in outcomes (the relatedness score) across a control (separate-topic pairs) and a treatment group (same-topic pairs). Since the relatedness measures produce metrics at different scales, it is not valid to compare raw differences in relatedness. An often-applied solution is to calculate Cohens D. However, this metric requires normal distributions for proper interpretation, and Table 1 and Figure 1 show that this assumption is clearly violated for all methods. Therefore, the non-parametric Cliff (1993) $\delta$ is calculated instead, which is defined as

$$\delta = \frac{\sum_{i,j} \mathrm{I}(x_i > x_j) - \mathrm{I}(x_i < x_j)}{n_{same} n_{seperate}} \qquad (9)$$

where the number of same- and separate-topic pairs are $n_{same}$ and $n_{seperate}$ with relatedness scores $x_i$ and $x_j$, respectively. In words, $\delta$ quantifies the proportion of same-topic pairs with higher relatedness than separate-topic pairs. It lies in the interval $[-1,1]$, with 1 (-1) signifying that every same-topic pair has higher (lower) relatedness than every separate-topic pair. In calculating $\delta$, we can only use publication pairs where both publications have received a relevance judgement to the same topic. 294,899 pairs meet this criterion, of which 22.5% are same-topic pairs.

**Benchmark test 2**

The second benchmark test evaluates the ability of the relatedness measures to correctly classify publications as belonging to a topic or not. We assume that a topic can be represented by a sample of publications that are judged relevant to that topic. If a relatedness measure is accurate, non-sampled publications also judged relevant should have a high relatedness score to the sampled publications.

Formally, the test does 50 iterations of the following for each topic. First, to define the sets $R_t$ and $NR_t$, which do and do not represent topic $t$, respectively, it samples 10 publications that are relevant and 10 that are not relevant to topic $t$. Let $p_i \notin (R_t \cup NR_t)$ be an unsampled publication that has received a judgement as either relevant or not relevant to topic $t$. Let $\max(\text{rel}(p_i, R_t))$ denote the maximum relatedness of $p_i$ to any $p_r \in R_t$, and let $\max(\text{rel}(p_i, NR_t))$ denote the maximum relatedness of $p_i$ to any $p_{nr} \in NR_t$. Then, the relevance judgement of $p_i$ to topic $t$ according to relatedness measure $r$ is defined as

$$rj_{itr} = \begin{cases} \text{relevant if} \max(\text{rel}_r(p_i, R_t)) > \max(\text{rel}_r(p_i, NR_t)) \\ \text{not relevant if} \max(\text{rel}_r(p_i, R_t)) \leq \max(\text{rel}_r(p_i, NR_t)) \end{cases} \quad (10)$$

That is, if relatedness measure $r$ shows that the publication most related to $p_i$ is in the sample of publications relevant to topic $t$, then $p_i$ is judged relevant to topic $t$ according to relatedness measure $r$.

The relevance judgements of the relatedness measures are calculated for each unsampled publication in each iteration, resulting in a total of 165,050 relevance judgements for each method. These are then compared to the experts' relevance judgements to calculate the number of true (T) and false (F) positives (P) and negatives (N), which are used to calculate Matthews (1975) correlation coefficient ($\varphi$), defined as

$$\varphi = \frac{TP \times TN - FP \times FN}{\sqrt{(TP + FP)(TP + FN)(TN + FP)(TN + FN)}} \quad (11)$$

In words, $\varphi$ shows the number of true positives and negatives relative to the number of false positives and negatives. A $\varphi$ of 1 (-1) means that every classification is true (false). We choose $\varphi$ over other popular alternatives for benchmarking binary classification systems because $\varphi$ includes all categories of the confusion matrix, which, e.g., the $F_1$-score does not.

# Results

We test three relatedness methods—Salton's cosine, soft cosine, and maximum term similarities—and evaluate their parameter settings using δ (test 1) and φ (test 2). Our parameter optimization tests: (1) major term weight values ($w$) across all measures, (2) IC weighting of term vectors for Salton's and soft cosine, (3) unweighted versus IC-weighted term distance graphs and λ values for soft cosine and maximum term similarities.

**Table 1** Parameter settings, accuracy, and descriptive statistics for most used version of Salton's cosine (first row) and different versions of three relatedness methods (other rows)

| Method | Parameters | | | | Performance | | Mean relatedness (skewness) for publications in… | |
| --- | --- | --- | --- | --- | --- | --- | --- | --- |
| | IC vector | Term distance | $w$ | $\lambda$ | $\delta$ | $\phi$ | Separate topics | Same topic |
| Salton's cosine | No | . | 1 | . | .328 | .173 | .118 (.892) | .178 (.711) |
| | No | . | 3 | . | .396 | .207 | .091 (1.640) | .186 (.734) |
| | Yes | . | 2 | . | .401 | .212 | .072 (1.847) | .156 (.946) |
| Soft cosine | No | 1 | 4 | 1 | .387 | .200 | .150 (1.273) | .256 (.535) |
| | No | ΔIC | 4 | 1 | .422 | .204 | .105 (1.632) | .211 (.726) |
| | Yes | 1 | 3 | 1 | .407 | .207 | .127 (1.462) | .229 (.694) |
| | Yes | ΔIC | 3 | 1 | .439 | .212 | .085 (1.877) | .186 (.927) |
| Maximum term similarities | . | 1 | 16 | 1 | .407 | .208 | .147 (1.199) | .259 (.506) |
| | . | ΔIC | 16 | 2 | .426 | .202 | .163 (1.131) | .282 (.467) |

Note: 294,899 publication pairs are used in computing Cliff's δ and the descriptive statistics. Of these, 66,208 are in the same topic, while 228,691 are in separate topics. 165,050 classifications are used in calculating ϕ. Skewness is calculated using Pearson's moment coefficient of skewness, $g_1 = \frac{\frac{1}{n}\sum_{i=1}^{n}(x_i-\bar{x})^3}{\left[\frac{1}{n}\sum_{i=1}^{n}(x_i-\bar{x})^2\right]^{3/2}}$, where $\bar{x}$ denotes the mean and $n$ is the number of observations.

**Optimal parameter setting**

Table 1 shows the optimal parameter settings for different versions of each relatedness method. Results are also shown for Salton's cosine without weighting terms by IC or weighting major terms more than minor ($w = 1$) because this is the most widespread approach. Results are only shown for Salton's cosine without qualifiers (**à**) because including qualifiers (**ǎ**) yields nearly identical results (see next section).

Five points emerge. First, all methods manage to compute higher mean relatedness to same-topic publication pairs than to separate-topic pairs. Second, the best-performing method is soft cosine with IC-weighted terms, major terms weighted by 3, and term similarities based on distances in the IC-weighted graph converted to similarity using $\lambda = 1$ (IC soft cosine $w3\ \lambda1$). Although Salton's cosine with IC-weighted terms and major terms weighted by 2 (IC Salton's cosine $w2$) yields a similar $\phi$, the $\delta$ of the soft cosine is 0.038 higher, making soft cosine's average score on the two tests higher. Third, the most used version of Salton's cosine (major and minor terms receive the same weight, terms not weighted by IC) has markedly lower accuracy than all other methods. In comparison to IC soft cosine $w3\ \lambda 1$, it has only 75% (82%) of the $\delta\ (\phi)$. Fourth, all versions of soft cosine achieve the best results with $\lambda = 1$.

The fifth point concerns the skewness of the methods, which is visualized in Figure 1. Compared to the most used version of Salton's cosine (standard), both IC Salton's cosine and IC soft cosine are more right skewed. This is likely because the most common term matches in a publication pair occur with the most common terms, which have low IC because of their commonness, and therefore these term matches contribute less to publication relatedness in the IC-weighted cosine similarities than in the conventional one. Of the methods with $\phi \geq .2$, the least right-skewed distributions are those of maximum term similarities. This indicates that this method might be better suited for distinguishing the relatedness between only distantly related publications. We return to this in the final section.

**Fig. 1** Density distribution of relatedness for publication pairs in separate and same topics for the most accurate version of Salton's cosine, soft cosine, and maximum term similarities, and for the most commonly used version of Salton's cosine

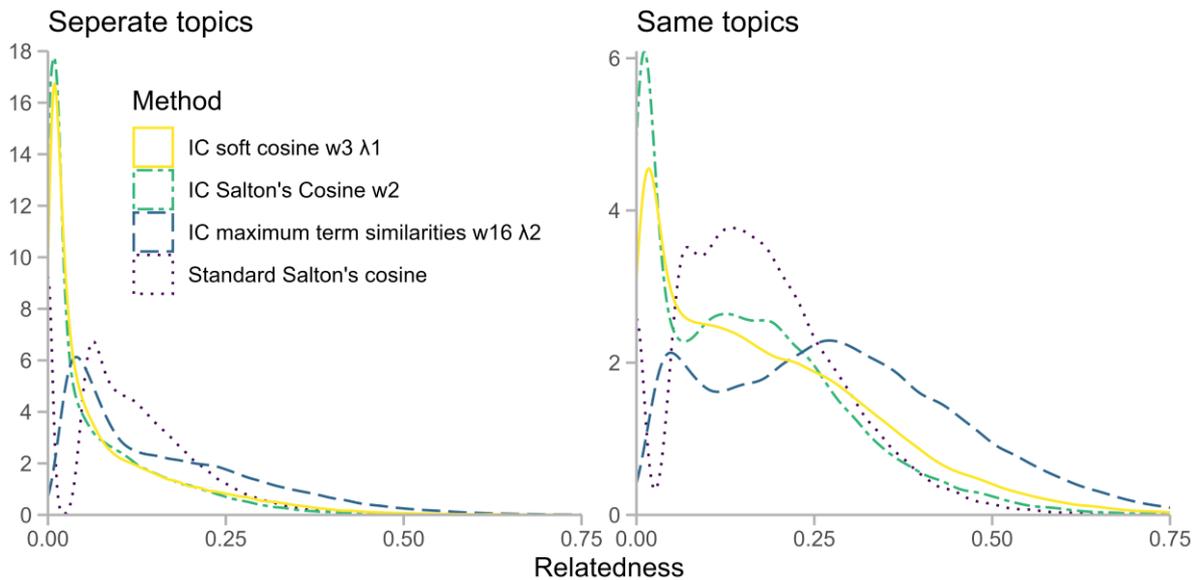

Note: The y-axes have different scales. X-axes truncated at 0.75; 0.2% (0.001%) of same-topic (separate-topic) pairs have relatedness>0.75. n same topic: 66,208; n separate topics: 228,691.

**Effects of parameters on accuracy**

We now inspect how the parameters affect the accuracy of Salton's cosine, soft cosine, and maximum term similarities. Figure 2 shows that including qualifiers in the term-vector representation of publications (**ă**) does not yield a substantively different accuracy than the simpler vector representation without qualifiers (**à**). This is likely because they produce very similar relatedness scores: Their concordance correlation coefficient (Lin, 1989)—a measure of agreement between measurements with 1 indicating perfect agreement—is 1 when rounded to five digits.[1]

**Fig. 2** Accuracy of Salton's cosine by major term weight and by IC weighting and qualifier inclusion. Darker tiles indicate higher accuracy

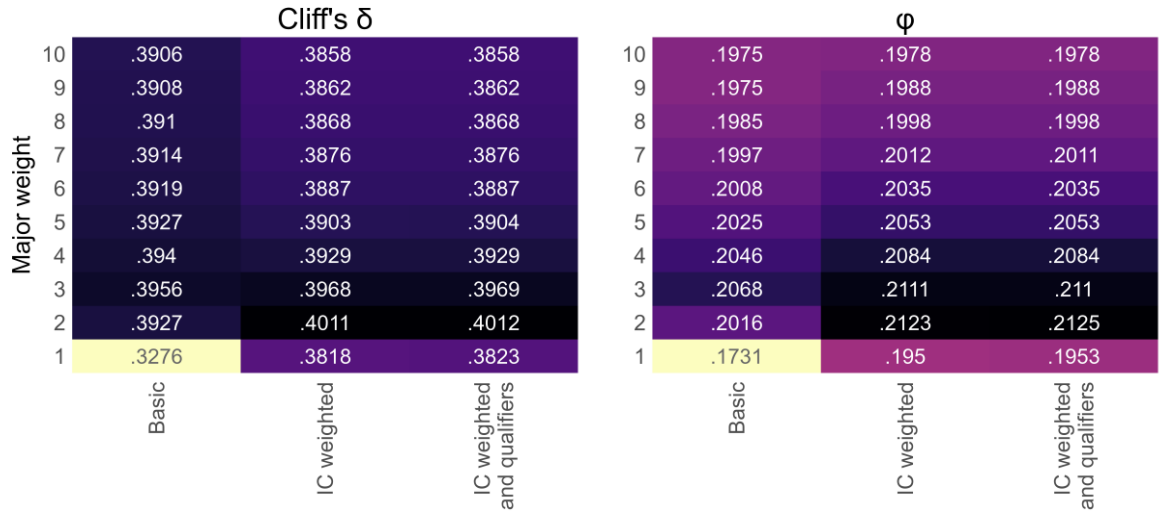

Figure 3 shows that increasing $\lambda$ decreases the accuracy of soft cosine regardless of the value of $w$. Comparing the first and second rows of Figure 4 shows that the "slim" approach, where minor terms are excluded, is more accurate than assigning major terms the same weight as minor terms. However, comparing the second row to any row above, i.e., comparing the "slim" approach to weighting major terms by any value between 2 and 20, shows that 'slim' is less accurate. Comparing the first column to any column to the left of it shows that converting distance to similarity using equation 4 yields higher accuracy than letting publication relatedness be a function of the minimum term distances.

## Discussion and conclusion

Quantitative science studies fundamentally depend on the accuracy of the measures they employ, and several well-developed literatures rest specifically on the validity of relatedness metrics (Hill et al., 2025; Lin & Wilbur, 2007; Myers, 2020; Stirling, 2007; Waltman & Van Eck, 2012). Inaccurate measures risk misclassifying the relationships between scientific publications, leading to flawed interpretations and misguided policy decisions. Improving the accuracy

---

[1] Rounded to eight digits, CCC ranges from 0.99996808 when $w = 1$ to 0.99999995 when $w = 10$. In comparison, the CCC between the IC-weighted and non-IC weighted vectors ranges from 0.77704304 when $w = 1$ to 0.95419296 when $w = 10$. $CCC = \rho \times C_b$, where $\rho$ is the Pearsons correlation coefficient, and

$$C_b = \frac{2}{\frac{\sigma_y}{\sigma_x} + \frac{\sigma_x}{\sigma_y} + \left(\frac{\mu_y - \mu_x}{\sqrt{\sigma_x \sigma_y}}\right)^2}$$

with $\mu_y - \mu_x$ accounting for differences in means and $\frac{\sigma_y}{\sigma_x} + \frac{\sigma_x}{\sigma_y}$ accounting for differences in standard deviation.

**Fig. 3** Accuracy of soft cosine by major term weight and $\lambda$ value used in term distance–similarity conversion. Darker tiles indicate higher accuracy

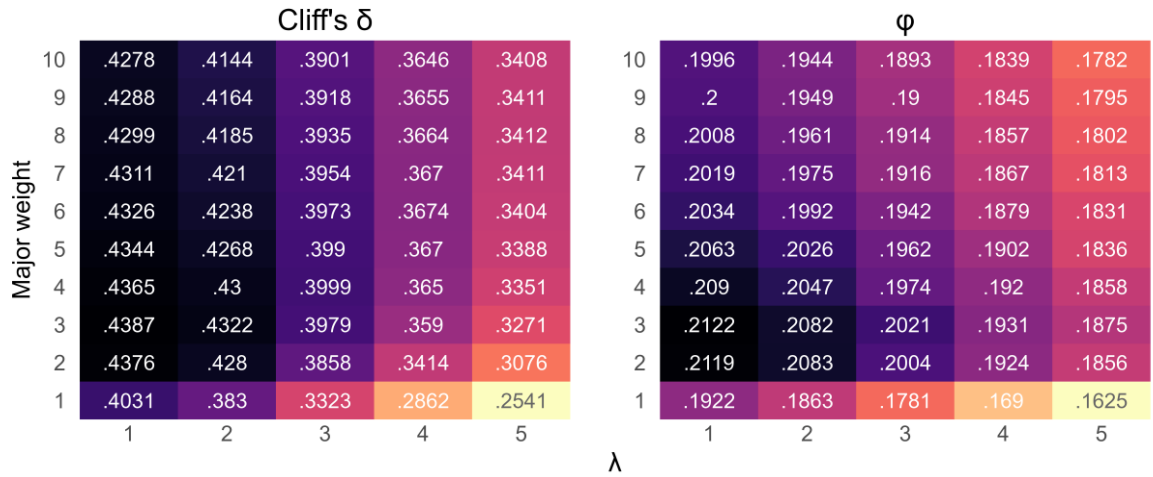

**Fig. 4** Accuracy of maximum term similarities by major term weight and $\lambda$ value used in term distance–similarity conversion is shown except in the bottom row and leftmost column. Bottom row: Minor terms are excluded (slim). Leftmost column: Term distances are used, and publication relatedness is a function of the minimum term distances. Darker tiles indicate higher accuracy

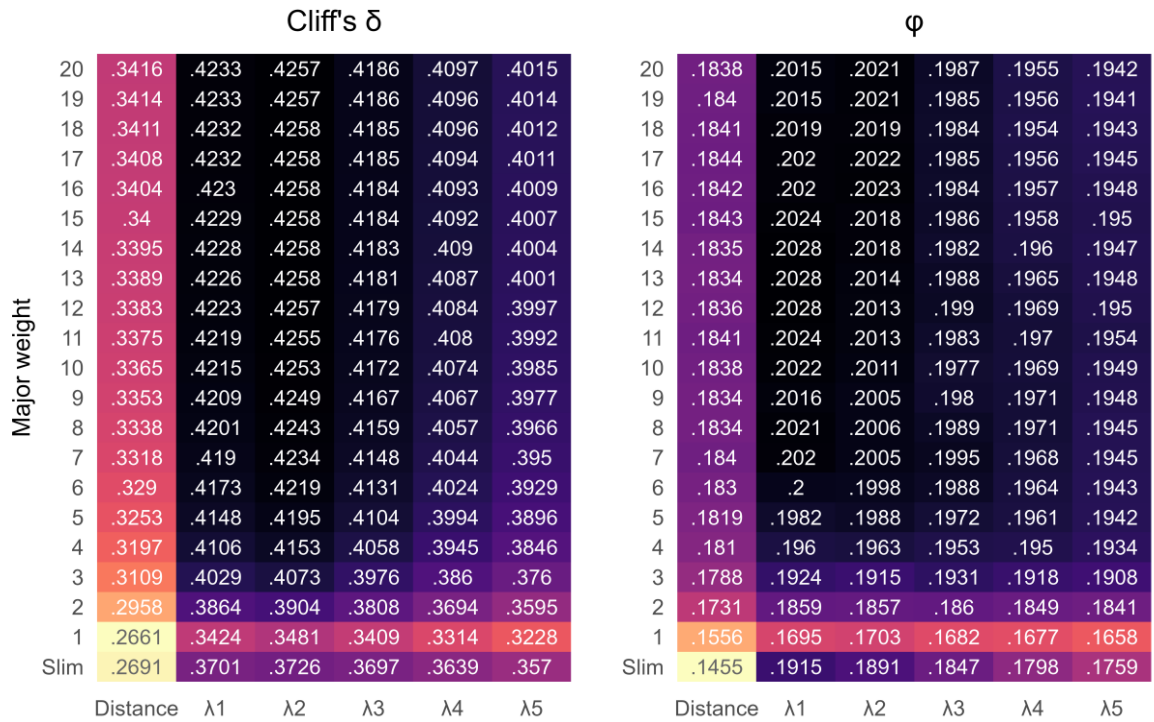

of relatedness measures is therefore not only a technical matter—it is central to the integrity and utility of quantitative science studies. Measures should, however, strike a balance between accuracy, feasibility, and parsimony. Choosing where to place that balance is ultimately a subjective decision that depends on the specific context and purpose of an analysis.

Currently, many studies simply use controlled vocabularies and Salton's cosine without IC weighting (Jia et al., 2017; Salatino et al., 2023; Yu et al., 2021) or—when using MeSH—weighting major terms more than minor terms (Boudreau et al., 2016; de Rassenfosse et al., 2022). The results in this article question whether these studies strike the right balance, since there are substantial gains in accuracy to be made by simply weighting by IC. The choice between IC soft cosine $w3\ \lambda1$ and IC Salton's cosine $w2$ is not as clear. Although the former has theoretical arguments in its favor, whether these and a 0.04 (9%) higher Cliff's $\delta$ justifies its greater conceptual and computational complexity may depend on its use case. Comparing Salton's cosine with and without qualifiers, the version with qualifiers is not worth the added complexity since the relatedness scores are nearly identical to those based on MeSH terms alone and offers no clear accuracy advantage.

Although maximum term similarities has slightly lower accuracy than soft cosine, its distribution is less right skewed. This indicates that the former is better at discerning the relatedness between distantly related publications. Where soft cosine clusters most relatedness scores near zero, maximum term similarities distribute them more evenly, enabling higher numerical differentiation between the relatedness of "very weakly related" and "weakly related" publication pairs. The utility of this feature depends on whether the analysis requires distinguishing between different levels of weak relatedness. In clustering and article recommender systems, it is not useful because the goals in these contexts are to cluster or recommend highly related publications. However, the feature could be useful in measuring interdisciplinarity across disciplines, the diversity of publication portfolios with disparate publications, or substantial changes to research trajectories.

In soft cosine and maximum term similarities, we used Zhu et al.'s (2009) formula (equation 4) in converting term distances to similarities, which requires researchers to specify a value of $\lambda$. In most cases, the best results were achieved with $\lambda = 1$. A simpler formula with less researcher degrees of freedom is therefore

$$\text{sim}(t_i, t_j) = \exp\left(-\text{dist}(t_i, t_j)\right) \quad (12)$$

A potential objection to the value of this article is that the proposed methods can only measure relatedness between publications indexed with a controlled vocabulary. Future research could use soft cosine or maximum term similarities as a large-scale benchmark to identify the most accurate relatedness method that does not rely on controlled vocabularies, similar to how Ahlgren et al. (2020) use Salton's cosine to benchmark relatedness measures in clustering.

The generalizability of this work is limited by only using MeSH as the controlled vocabulary and TREC Genomics as the benchmark data. Future research could replicate and extend the tests using other controlled vocabularies and benchmark datasets. For example, Cochrane reviews could serve as a benchmark for MeSH under the assumption that publications included in the same review are related. Alternatively, "the Integrated Search Test Collection", which contains relevance judgements of documents to search tasks in physics (Lykke et al., 2010), could be used to benchmark methods based on Physics Subject Headings (PhySH) or the Physics and Astronomy Classification Scheme (PACS).

**Acknowledgments** A previous version of this article was presented at the 28th International Conference on Science, Technology and Innovation Indicators 2024 (Alnor, 2024). The present article incorporates substantial revisions, notably the addition of soft cosine, the distance–similarity conversion, and tests of major weights greater than 3. The author thanks Jens Peter Andersen, Vincent Traag, the conference participants, and colleagues at the Danish Centre for Studies in Research and Research Policy for their comments and suggestions.

**Data and code availability** The data used in the article is publicly available from Oregon Health & Science University (relevance judgements) and the National Center for Biotechnology Information (MeSH-terms). The article uses the open-source statistical software R, and the script is available at https://github.com/EDAlnor/MeSH-relatedness. All results can be replicated by running the script, which will download, process, summarize and visualize the data.

**Conflict of interest** The author has no conflict of interest to declare that are relevant to the content of this article.

**Funding information** No funding was received for this study.